\documentclass[12pt]{article}
\usepackage{amsmath}
\usepackage{amsthm}
\usepackage{appendix}
\usepackage{graphicx}
\usepackage{verbatim}
\usepackage{epsfig}
\usepackage{hyperref}
\usepackage{float}
\usepackage{color}
\usepackage{csquotes}
\usepackage{fullpage}
\usepackage{amsfonts}
\usepackage{amscd}
\usepackage{amssymb}
\usepackage{graphics}
\usepackage{amsmath}
\usepackage[english]{babel}
\usepackage{bbm}
\usepackage{yfonts}
\usepackage{mathrsfs}
\usepackage{color}
\usepackage{cleveref}
\DeclareFontFamily{OML}{rsfs}{\skewchar\font'177}
\DeclareFontShape{OML}{rsfs}{m}{n}{ <5> <6> rsfs5 <7> <8> <9>
rsfs7 <10> <10.95> <12> <14.4> <17.28> <20.74> <24.88> rsfs10 }{}
\DeclareMathAlphabet{\mathfs}{OML}{rsfs}{m}{n}

\newcommand{\bae}{\begin{equation}\begin{aligned}}
\newcommand{\eae}{\end{aligned}\end{equation}}

\theoremstyle{definition}
\newtheorem{example}{Example}

\appendixtitleon
\appendixtitletocon

\begin{document}
\numberwithin{equation}{section} \numberwithin{figure}{section}
\title{A Bayesian framework for analyzing alleged cheating in sports through hidden codes, with applications to bridge and baseball}
\author{Aafko Boonstra\footnote{Department of Mathematics, Vrije Universiteit Amsterdam; a.j.w.boonstra@vu.nl.} \, and Ronald Meester\footnote{Department of Mathematics, Vrije Universiteit Amsterdam; r.w.j.meester@vu.nl.}}
\maketitle

\begin{abstract}
We develop a statistical framework to evaluate evidence of alleged cheating involving illegal signaling in sports from a forensic perspective. We explain why, instead of a frequentist procedure, a Bayesian approach is called for. We apply this framework to cases of alleged cheating in professional bridge and professional baseball. The diversity of these applications illustrates the generality of the method.
 \end{abstract}

\section{Introduction}
\label{intro}
The European Commission acknowledged the essential role of sport in European society in their 2007 white paper \cite{ec}, highlighting the integrity of sports as one of the priorities of EU policy. Clearly, any form of cheating, in any sport, is incompatible with sports integrity. In various disciplines and sports, certain types of illegal signaling could provide a significant advantage. For instance, in one case in bridge, the orientation of the lead card was allegedly used to transfer useful information about the hand. We find another example in baseball, where sound signals were allegedly used to transfer information concerning the type of ball the pitcher was about to throw. In this paper, we present a Bayesian statistical analysis that adequately evaluates the evidence in such cases of alleged cheating. 

Statistics are routinely used in order to detect certain types of cheating \cite{reyhan} or match fixing \cite{forrest} in various sports. In most cases, statistical inference provides leads or serves as a starting point of an investigation rather than being part of the evidence. For example, \cite{robinson} used extreme value techniques to find inconsistencies in the data of annual best times for the women's 3000m track event. Such findings are generally not sufficient to draw conclusions on any hypothesis of cheating, but they may provide ground to conduct additional doping tests.

Frequentist methods, such as the well-known hypothesis testing procedure and the use of $p$-values, are widely used to assess the evidential value of hypotheses. However, it has been recognized by many scholars that such frequentist methods are not satisfactory, for a number of reasons. (See for instance \cite{briggs, meesterslootenbook, fluke, royall} and the many further references therein.) The key issue is that a frequentist method yields answers that predict how (un)likely the data are under the assumption of innocence. This, however, is an answer to the wrong question. It is not interesting at all how unlikely the data is under a given hypothesis. The real question is that given the data, which hypothesis is most likely responsible for it? The fact that the data were unlikely from the outset is no longer important once the data are there. They have to be explained and the question is which hypothesis explains the data best. 

This question leads to the consideration of multiple candidate hypotheses, which is standard in forensic matters. One formulates at least two hypotheses and then tries to evaluate how well these hypotheses explain the data. If the first hypothesis explains the data (much) better than the second (i.e., the probability of the data under the first hypothesis is (much) higher than under the second), the data serve as evidence for the first hypothesis relative to the second, and vice versa. The ratio of these probabilities is called the {\em likelihood ratio}. In our context, the two hypotheses of interest are, firstly, the hypothesis that the defendants used a specified (illegal) method to transmit information, and secondly, the hypothesis that they did not. This approach is Bayesian in nature. 

In this article, we present a Bayesian analysis to assess the statistical evidence in two cases of alleged cheating. The first is the Fantoni-Nunes vs.\ EBL case in bridge, where the allegation was that the physical orientation of certain cards revealed information about the hand. Most consulted experts at the time applied some form of frequentist hypothesis testing (notably \cite{lawler, buchen}), although some reports do contain some Bayesian elements (\cite{regan, viret}). These reports are available as supplementary material to this paper, with permission from the EBL. We will argue that instead, a full Bayesian analysis is more appropriate and called for. 

A similar case occurred in the Major League Baseball (MLB) with the Houston Astros cheating scandal in 2017, where the allegation was that (illegal) information about the pitch to be thrown was transmitted by banging a trash can.

Although baseball and contract bridge are very different, the cases were surprisingly similar and can be treated with virtually the same mathematical model.
We will provide all details and necessary context of these two cases below. It will be clear from the way we formulate our models that our approach should be suitable in many more specific circumstances, and in possibly many different disciplines.

The paper is structured as follows. In Section \ref{bayes}, we will first provide some necessary background on Bayesian reasoning and its general importance for forensic science and statistics. 

We then turn to the bridge case. For readers unfamiliar with bridge, we discuss (in Section \ref{char}) the basics of contract bridge and provide more details in the appendices. In Section \ref{analysis}, we develop and discuss a statistical framework to evaluate the evidence in the aforementioned case. In Section \ref{honkbal}, we explain how this framework applies to the above mentioned cheating scandal in professional baseball. We end with a discussion and some conclusions in Section \ref{conclusions}.

\section{Bayesian reasoning}
\label{bayes}
Before we go into the actual cases, we support our conviction that a Bayesian approach is called for by quoting from the ENFSI guidelines, written under the auspices of the European Network of Forensic Science Institutes \cite{enfsi}. The aim of the guidelines was to ``standardize and improve evaluative reporting in ENFSI laboratories''. The project was undertaken by a ``core group of scientists from member institutes''. Many of the recommendations in the document concern the use of probability and statistics.

The report is unambiguous about the use of the likelihood ratio and the fact that classical statistical methods (like $p$-values)  should not be used for evidential purposes. On page 6, we read: 

\begin{quote}
``Evaluation [...] is based on the assignment of a likelihood
ratio. Reporting practice should conform to these logical principles.
This framework for evaluative reporting applies to all forensic science
disciplines. The likelihood ratio measures the strength of support the
findings provide to discriminate between propositions of interest. It is
scientifically accepted, providing a logically defensible way to deal with
inferential reasoning. Other methods (e.g., chemometrical methods)
have a place in forensic science, to help answer other questions at
different points of the forensic process (e.g., validation of analytical
methods, classification/discrimination of substances for investigative
or technical reporting). Equally, other methods (e.g., Student's t-test)
may contribute to evaluative reports, but they should be used only
to characterize the findings and not to assess their strength. Forensic
findings as such need to be distinguished from their evaluation in the
context of the case. For the latter evaluative part only a likelihood ratio
based approach is considered.''
\end{quote}

This is an important principle. It emphasizes that only a likelihood ratio  can measure the strength of the evidence, and that other statistical methods are not able to do that. The principle implicitly assumes that one can only meaningfully speak about evidential value in a comparative way. This is made explicit on page 10 of the report, where we read:  

\begin{quote}
``The findings should be evaluated given at least one pair of
propositions: usually one based upon one party's account of the events and
one based upon an alternative (opposing party's) account of the events. If no
alternative can be formulated, the value of the findings cannot be assessed. In
that case, forensic practitioners should state clearly that they are not reporting
upon the value of the findings.''
\end{quote}

A similar statement can be found on page 13:

\begin{quote}
``The report should stress that in the absence of an alternative proposition, it is
impossible to evaluate the findings.''
\end{quote}

The report also discusses the way the forensic expert should formulate his or her findings. On page 10 we read:

\begin{quote}
``The conclusion shall be expressed either by a value of the likelihood
ratio and/or using a verbal scale related to the value of the likelihood
ratio. The verbal equivalents shall express a degree of support for
one of the propositions relative to the alternative. The choice of the
reported verbal equivalent is based on the likelihood ratio and not
the reverse. The report shall contain an indication of the order of
magnitude of the likelihood ratio.''
\end{quote}

and

\begin{quote}
``Note that if a likelihood ratio cannot be assigned by the forensic practitioner
(due to a lack of knowledge for example), then no appropriate evaluative
assessment of the findings can be made.''
\end{quote}

Basically, the report says that the likelihood ratio framework should always be used regardless of available data, and that if it does not allow to give a result, then no evaluation is possible. This is, in fact, tantamount to what was said previously, namely that the likelihood ratio framework is the only acceptable one. 

Finally, the ENFSI report contains a philosophical statement about the use of probabilities. On page 23 we read:

\begin{quote}
``Your subjective probability  is the measure for your belief in the occurrence
of an event. A number between 0 and 1 represents this measure. The laws
of probability apply to these probabilities just as they apply to calculated
probabilities.
A measure of belief might be obtained by doing thought experiments, and
possibly further informed by ad hoc small-scale physical experiments. Expert
knowledge elicitation is a more technical approach to obtaining subjective
probabilities.''
\end{quote}

Probability theory plays a very different role in court compared to fields such as operations research or financial mathematics, because the goal is not to predict the likelihood of certain events. In legal cases, the events of interest have already occurred. The suspect is guilty or innocent, and any probabilistic statement will only reflect the conviction of the judge or expert. Probabilities in law do not predict, but aim to express which hypothesis provides the best explanation for the observed events. A probability in law is therefore an {\em epistemic} probability, a measure to quantify the degree of belief in a specific hypothesis. Epistemic probabilities are also {\em personal}, in the sense that different individuals may arrive at different levels of belief, based on a difference in (the interpretation of) information. Here, the phrase ``personal'' still means that a personal probabilistic statement requires a solid justification. 

Bayesian reasoning essentially consists of formalizing the distinction between belief and evidence by distinguishing between context and evidence. In the formal Bayesian approach, the evidence is summarized by the likelihood ratio that we discussed above. 
As an example, $H$ could be the hypothesis that an athlete $A$ used doping to enhance their performance, while the alternative hypothesis $H'$ is that no illegal substances were used. Within this framework, we already mentioned that we want to be able to speak about the probability of hypotheses  $H$ and $H'$.  It can be someone's informal personal conviction that $H$ is true, and it may also express the degree to which a judge or juror is convinced by $H$. In any case, we are interested in the probability of $H$ or $H'$ {\em given the evidence}.

Since the hypotheses are examined in relation to one another, it is convenient to speak of the {\em ratio} between the probabilities of $H$ and $H'$ upon observing the evidence. This ratio is called the {\em posterior odds}.  The context tells us how we should see the probability ratio {\em prior} to observing the evidence. This ratio is called the {\em prior odds}. In the example, this is the probability that $A$ cheated with doping relative to the probability that $A$ participated according to the rules. The intuitive fact that one's final conviction depends on both context (the prior) and the evidence (the likelihood ratio) is conveniently summarized in the mathematical statement that
$$
\mbox{ posterior odds } = \mbox{ likelihood ratio } \times \mbox{ prior odds},
$$
which is nothing but the well know Bayes' rule from elementary probability theory.

The distinction between the context and the evidence is mathematically very useful, but to some extent arbitrary. Whether someone counts a fact as a `circumstance' or as `evidence' ultimately does not matter, but usually there is a natural choice. For example, in the bridge case we view the fact that playing cards were laid down both horizontally and vertically in our statistical analysis as background information. Legally, this fact is part of the evidence. 

The presumption of innocence finds its way into this paradigm by insisting that the prior probability that someone is guilty must be very small. Not zero $\--$ otherwise a person would not stand to trial. In addition, uncertainty will always be dealt with in the advantage of the defendant. We refer to \cite{meesterstevens} for an in-depth discussion of this issue in relation to Bayesian reasoning.

\section{The bridge case}

For readers who are unfamiliar with contract bridge, we outline the fundamental principles of the game. For more information, including a discussion on the impact of cheating and a comparison with other mind sports such as chess, please see Appendix A.

\subsection{The basics of contract bridge}
\label{char}

The fundamental ingredients of the game are a set of four players with a standard deck of 52 cards. The four players are indicated by the cardinal directions North, East, South, and West, and are seated accordingly. Every player receives 13 cards, distributed at random. The four players do not play individually but work in fixed teams of two players, usually referred to as \textit{partnerships} or \textit{pairs}. The pair North-South (NS) plays against the pair East-West (EW). During play, a player reveals one card from their hand, followed by the others in a clockwise rotation. Each such set of four cards is called a \textit{trick}, so that a single game or \textit{board} consists of 13 tricks. A trick is won by the player who contributed the card with the highest value. This player will then lead off a card to start the next trick, and this procedure continues until all cards are played. After 13 tricks, both partnerships add up the tricks their respective players have collected. The goal for the partnership, and thus for the players, is to collectively win as many tricks as possible.

The phase of playing the 13 consecutive tricks is preceded by a phase called the \textit{auction}. The auction results in a \textit{contract} for one of the pairs, say NS. The other pair, EW, now has to \textit{defend} this contract by playing the first card of the board. This card is referred to as the \textit{lead}. The contract prescribes how many tricks NS should take at least to gain a positive score. If NS fails to gain enough tricks, the board results in a positive score for EW. Furthermore, a contract also determines whether there will be a special suit, the \textit{trump} suit. If so, the contract specifies this suit (either spades, hearts, diamonds, or clubs).

A special feature of bridge is that if a pair plays a contract, one member of this pair reveals their hand to both their partner and the opponents for the rest of the board. This player is called the \textit{dummy} and is not actively involved in the play anymore. His partner, the \textit{declarer}, decides which card is played from the dummy’s hand.

The case in this paper is concerned with the leads. It is common for partnerships to establish agreements about their leads. One such agreement involves leading the lowest card from a suit containing an \textit{honor} (e.g., the ace, king, queen, or jack, and sometimes the ten), while leading the second-lowest card from a suit with no honors. For instance, a partnership with this understanding would lead the 2 from a hand like A9762, and the 7 from a holding like 9762. This information is valuable to defenders, as it can improve their decision-making as the play progresses.

Furthermore, an agreement like code C has the added benefit that declarer is unaware of the signaling, since tournament bridge requires all basic conventions to be disclosed before a match, ensuring both teams are informed. In the given example, if the 2 is led, both the declarer and the other defender will understand that it indicates an honor is held in the suit. During the match, defenders must disclose their signaling methods if requested. While signaling is permitted and plays a vital role in the game, it must be done through the spots on the played cards. Physical signaling, such as sounds, gestures, or card orientation, is strictly prohibited, whether disclosed or not.

\subsection{A Bayesian analysis}
\label{analysis}

It is uncontested that Fantoni and Nunes (FN) vary in the way they lay down their leads when they defend. In most cases, the leads can be identified as either clearly horizontally or clearly vertically placed. We will therefore assume that cards can always be classified as either horizontally or vertically placed, neglecting the unlikely event of an ambiguous diagonal card. The alleged code in the court case was as follows:

\begin{quote}
    If a card is led horizontally, this denies a top honor (ace, king or queen) in the lead suit. Conversely, if the card is lead vertically, this implies holding a top honor in that suit. This rule does (naturally) not apply in case the lead card was a singleton, that is, the lead card was the only card in that suit.
\end{quote}

We will refer to this code as \textit{code C}.

\subsubsection{The hypotheses}

Any Bayesian analysis starts with identifying two or more competing hypotheses. We formulate two hypotheses denoted $H_C$ and $H_R$ respectively: 

\medskip
$H_C$: FN used code $C$;

 $H_R$: The orientations of the leads were random and independent of the cards held. 

\medskip\noindent
Note that the two hypotheses are not each others complement. 

Next, we identify the data. In board $i$ we write $X_i$ for the 13 cards in the hands of the player playing the lead, and $L_i$ for the lead played. For every $i$ we also have the orientation $O_i \in \{0,1\}$, labeling the direction of the lead. We let $O_i=1$ correspond to a horizontal lead in the $i$th board and $O_i=0$ to a vertical one. Assuming that we play $n$ boards, we write $\mathbf{O}:=(O_1, \ldots, O_n)$ and
$$
\sum_{i=1}^n O_i = h
$$
for the observed number of horizontal leads.

The alleged code $C$ determining the orientation of every lead depends on both $X_i$ and $L_i$ and results in either a 0 (vertical) or a 1 (horizontal): 
$$
C: (X_i, L_i) \to \{0,1\}.
$$ 

For every $i$, we can now determine the \textit{match indicator} $M_i$ as follows:

\[ M_i = 
\begin{cases}
1, \text{ if } C(X_i, L_i)=O_i; \\
0, \text{ otherwise.}
\end{cases} 
\]
That is, we have a match ($M_i =1$) if the actual orientation corresponds correctly to what the code $C$ predicts in the $i$-th board. Finally we write 
$$
\sum_{i=1}^n M_i =m
$$
for the observed number of matches.

\subsubsection{The models}
The next issue on the agenda is to formulate suitable models under both hypotheses. Indeed, if we want to be able to speak about the probability of the data under each hypothesis, we need a mathematical model in which such probabilities make sense and can be calculated. This modeling is crucial, of course. It should be in some sense `close' to reality, but at the same time it should be simple enough to allow computations. We take $\mathbf{O}$ as our principal data (but we will in fact use only $h$ and $m$, which are derived from $\mathbf{O}$), and use the other information as background needed to compute probabilities.

We therefore need a model for $\mathbf{O}$, under each hypothesis. Under $H_C$, the player follows the code $C$. If this were done flawlessly, then the probability of the observed $\mathbf{O}$ would be either 0 or 1, depending on whether $O_i = C(X_i, L_i)$ for every $i$ or not. However, we will allow for the possibility of making mistakes, and suggest a model in which every lead corresponds to the code with some (high) probability $p$. We assume that the code is time-homogeneous. If we also assume independence between different boards, the probability of the observed orientations is then equal to
$$
P(\mathbf{O} \mid H_{C}) =  p^m (1-p)^{n-m}.
$$
Note that perfect coding $p=1$ indeed results in a probability of either 0 or 1. 

Under $H_R,$ the observed orientations are random and independent of the cards held. This implies that a possible match is purely coincidental. Although it is undeniable that different players may have different, personal ways of physically laying down the lead cards, we think that it is reasonable to assume that both players have the same, fixed time-homogeneous probability $q$ of placing the card horizontally and a fixed probability $1-q$ of playing it vertically, allowing for different values of $q$. If we do not want to favor certain values of $q$ over other values, we have to take a uniform distribution for this $q$.  This leads to
$$
P(\mathbf{O} \mid H_R)= \int_0^1 q^h(1-q)^{n-h} dq.
$$

\subsubsection{The likelihood ratio}
We have now finalized the set-up, and we can perform computations for various values of $n, m, h$ and $p$. In the FN-case under consideration, we have $n=85$, $m=83$ and $h=45$ \cite{casuitspraak}. If we choose $p=0.9$, we obtain
\begin{equation}
\label{LRone}
LR(\mathbf{O}) = \frac{P(\mathbf{O} \mid H_{C})}{P(\mathbf{O} \mid H_R)} \approx 4 \times 10^{19}. 
\end{equation}

In words, if the players follow the code $C$ with probability $p=0.9$, then the actual outcome $\mathbf{O}$ is about $4 \times 10^{19}$ times more likely under $H_C$ than under $H_R$. This likelihood ratio is, therefore, extremely high, providing extremely strong support for $H_C$. Choosing higher values for $p$ will result in even higher likelihood ratios.

It is important to keep in mind that a high likelihood ratio in favor of $H_C$ only will generally not guarantee a high posterior probability of $H_C.$ This common pitfall of excluding the prior odds is known as the \textit{prosecutor's fallacy} (\cite{meesterslootenbook} provides many real-life examples). Both the likelihood ratio and the prior odds are calculated based on models and their subsequent choices of parameters. The possibles consequences of these choices for the posterior odds will be discussed in the next section.

\subsubsection{Prior and posterior odds}
The task of an expert witness in a court case is in principle restricted to evaluating the evidence by means of a likelihood ratio. But as we explained, in order to use this likelihood ratio, one should combine this with a suitable prior. In this subsection we make some observations that can guide a legal decision maker to this end, but we remark that the decision about a prior is not ours to make. 

Actually, in this case, the likelihood ratio is so huge that it will probably convince anyone, independent of one's prior ideas about the probability of cheating. (Except, of course, if your prior probability of cheating is 0; in that case no evidence can ever change your mind.)

So, let $\psi$ denote one's prior probability that FN cheated using the orientation of the leads. Obviously opinions will vary concerning the value of $\psi$ but it should not be 0. This prior probability of cheating is not equal to  $P(H_C)$: indeed one could in principle cheat with different codes. However, the number of such codes is clearly limited, say by the number $M$. We stress that the choice of $\psi$ and $M$ is personal, but at the same time it is not so that anything goes: any choice should be argued for. See Appendix B for a small discussion concerning the value of $M$.

If we assume that {\em if} a certain pair cheats, the probability to use code $C$ is at least $1/M$, {\em then} the prior odds are given by
$$
\frac{P(H_C)}{P(H_R)} = \frac{\psi \times 1/M}{1-\psi}.
$$

Combining this with the likelihood ratio \eqref{LRone} gives
\begin{equation}
	\begin{split}
	\text{posterior odds} &= \text{likelihood ratio} \times \text{prior odds} \\
						&\approx \frac{4 \times 10^{19}}{M} \times \frac{\psi}{1-\psi} 			
	\end{split}
\end{equation}

The reader may at this point wonder why we did not address the `freshness' of the data, an issue raised by several experts, e.g.\ \cite{lawler}. It was argued that the data used to formulate an allegation cannot also be used as evidence. This in order to prevent unjustified confidence in what is called a {\em data-driven hypothesis}. 

However, Bayesian reasoning naturally deals with such situations, because ill-constructed hypotheses will simply have a very small prior probability. Even if a pair randomly alternates between leading horizontally or vertically, there will almost always be some code that fits the data very well, but this code would probably not be considered useful at all. There are not so many useful codes (see Appendix B), so that a code fitting a random outcome of directions would typically have an extremely small prior probability, implying that its posterior probability is still small, despite the high likelihood ratio. 

All these issues are well known in forensic probability and statistics - see e.g.\ Example 2.2.1 in \cite{meesterslootenbook}, where a large likelihood ratio in favor of a data-driven hypothesis does not lead to high posterior odds since the prior odds are so small.

\section{The baseball case}
\label{honkbal}
The situation here is, as we will see below, very similar to the bridge case. Therefore, our description in this section can be somewhat more concise than in the bridge case. In addition, the baseball rules are probably much better known in general than the bridge rules.

\subsection{The case}
The baseball case started in 2019 with an article on the subscription-based sports news website \textit{The Athletic} reporting electronic `sign stealing' by the Astros \cite{rosenthal}. In \cite{stitzel}, the relevance of sign stealing is described as follows:

\begin{quote}
    ``The focus of baseball is the competition between the batter and the pitcher. The pitcher seeks to get the batter “out” by causing him to miss three pitches or by inducing weak contact enabling the defenders to retire the batter on the basepaths. To obtain an out, the pitcher uses various pitch types and pitch locations to disrupt the batter’s ability to hit the pitch. The batter’s uncertainty about the pitcher’s intentions on any given pitch is paramount to the pitcher’s success. The pitcher and his catcher, communicate via the catcher’s hand signals. These hand signals tell the pitcher which pitch type and where to throw.

    [...]

    The Astros positioned video cameras in center field of their home stadium and trained them to steal the opponent’s catcher’s signs to the pitcher. They then conveyed this information in real-time to an operator behind their team’s dugout who used decoding software and audible signals to communicate the pitch information to the batter in time for him to react to the pitch.''
\end{quote}

Videos highlighting the noises went viral \cite{spackman}, and eventually led to an investigative report by the MLB \cite{manfred}. Concerning these audible signals, the report states the following:

\begin{quote}
    ``One or more players watched the live feed of the center field camera on the monitor, and after decoding the sign, a player would bang a nearby trash can with a bat to communicate the upcoming pitch type to the batter. (Witnesses explained that they initially experimented with communicating sign information by clapping, whistling, or yelling, but that they eventually determined that banging a trash can was the preferred method of communication.) Players occasionally also used a massage gun to bang the trash can. Generally, one or two bangs corresponded to certain off-speed pitches, while no bang corresponded to a fastball.''
\end{quote}

According to \cite{stitzel}, the off-speed pitches fall into 10 categories such as a `curveball', `slider', `changeup' etc. As in the Fantoni-Nunes vs.\ EBL case, possible codes were quickly constructed based on the videos. Contrary to the bridge case, the investigation committee had additional evidence, including 68 witnesses and ``tens of thousands of emails, Slack communications, text messages, video clips, and photographs'' \cite{manfred}. 

In 2020, the Astros fan Tony Adams compiled a large database \cite{adamsdata} of home games played in Houston. The data records sounds of banging for over 8200 pitches together with the type of pitch. Most of the research with this data seems to address the question whether the cheating improved the performance, but we also found a (frequentist) analysis focusing on cheating \cite{olbrecht}. Adams shared the data for others to analyze it, but he did propose a particular code himself. We found very strong evidence for a slightly different but also very simple code, that we refer to as $code$ $B$:

\medskip\noindent
{\bf (Code B)}
If the catcher gives a sign for a fastball, that is, a four-seam fastball, a two-seam fastball, a cutter, a sinker or a split-finger fastball, no trash can is banged. Vice versa, if the catcher gives a sign for any non-fastball, such as a breaking ball or a changeup, the trash can is banged.

\subsection{A Bayesian analysis}
Following our approach in the bridge case, we write down the following hypotheses:

\medskip
$H_B$: The Astros used code $B$;

$H_R$: The sounds of banged trash cans occurred independently of the pitches of the opposing team. 

\medskip
As before, the two hypotheses are not each other's complement. Let $X_i$ denote the type of pitch $i$. For every $i$, $S_i \in \{0,1\}$ denotes the  occurrence of a bang ($S_i=1$) or the absence ($S_i=0$). Denoting the number of observed pitches by $n$, we write $\mathbf{S}:=(S_1, \ldots, S_n)$ and
$$
\sum_{i=1}^n S_i = b
$$
for the observed number of banged trash cans. Note that $\mathbf{S}$ plays the role of $\mathbf{O}$ in the bridge case. 

The alleged code $B$ determining the sound accompanying every pitch depends on $X_i$ and results in either a 0 (no bang) or a 1 (a bang): 
$$
B: X_i \to \{0,1\}.
$$ 

For every $i$, we can now determine the \textit{match indicator} $M_i$ as follows:

\[ M_i = 
\begin{cases}
1, \text{ if } B(X_i)=S_i; \\
0, \text{ otherwise.}
\end{cases} 
\]
That is, we have a match ($M_i =1$) if the actual sound corresponds correctly to what the code $B$ predicts for the $i$-th pitch. Finally we write 
$$
\sum_{i=1}^n M_i =m
$$
for the observed number of matches.

As before, we model the scenario under $H_B$ using a parameter $p$ that represents to what extent the team is able to successfully carry out the code. This value should realistically be lower compared to the bridge case, because the road from the sign of the catcher to a correct sound registration in a data file is generally more bumpy. For example, the sign of the catcher needs to be interpreted and transmitted correctly within only a handful of seconds. Apart from that, the opposing pitcher will also need to deliver the throw the catcher signaled for, and on top of that this throw also has to be registered correctly. As before, we write
$$
P(\mathbf{S} \mid H_{B}) =  p^m (1-p)^{n-m}.
$$
For $H_R,$ we again propose a binomial model, but here there seems to be no need to integrate over all values between 0 and 1. That is because in the period between April 3 and May 24th, almost no bangs were observed, so we can be quite sure that no cheating with bangs happened during this period. Hence, we can use the data over that period to get an idea of reasonable values for the probability $q$ of a bang under normal circumstances. During the 22 home games that were played during that period, the average number of bangs per pitch is 0.015. Viewed per game, this average peaked at 0.03. It seems, therefore, quite safe to bound $q$ from above by 0.1:

$$
P(\mathbf{S} \mid H_R)= \int_0^{0.1} q^b(1-q)^{n-b} dq.
$$

We are now in a position to perform computations for various values of $n, m, b$ and $p$. For example, in the matches of June 31 and July 1 against the New York Yankees, we have $n=267$, $m=201$ and $b=85$. If we choose $p=0.8$, we obtain
\begin{equation}
\label{LR1}
LR(\mathbf{S}) = \frac{P(\mathbf{S} \mid H_{B})}{P(\mathbf{S} \mid H_R)} \approx 3.4 \times 10^{30}. 
\end{equation}
Thus, if the team follows the code $B$ with a probability of $p=0.8$, then the actual outcome $\mathbf{S}$ is 3.4 \textit{nonillion} times more likely under $H_B$ than under $H_R$. We found high values for the likelihood ratio in most of the home games played between May 28 and September 21. Aggregating over all of the data leads to large numbers that we couldn't easily process, but the matches against e.g. Minnesota Twins (July 15, 16), Toronto Blue Jays (August 4, 5, 6) and Chicago White Sox (September 19, 20, 21) yield likelihood ratios of orders $10^{9}, 10^{23}$ and $10^{12}$ respectively. Very similar to the bridge case, the possible ways to cheat meaningfully with banging are limited, mitigating the data-driven nature of $H_B.$ 

Although the data yields extreme likelihood ratios $\--$ partly due to the sheer size of the dataset $\--$ it still overlooks the most important factor that made the cheating so obvious to viewers. Indeed, the data does not say anything about the \textit{timing} of the bangs $\--$ we simply did not take this into consideration yet. Almost all bangs occur after the catcher's signaling and just before the pitcher begins the pitch. If the data consisted of time frames, say 30 seconds long, that include both the signaling and the pitch, most of the registered bangs would fall within this time window. The likelihood of that event is close to 1 under \textit{any} hypothesis of cheating with bangs, while it rapidly decreases if no relationship between these events is assumed. The pre-designated time windows are usually not larger than about 6 seconds, which in this hypothetical scenario would result in a factor of at least 5 for every bang towards the hypothesis of cheating. If we would take this into account in our modeling, the likelihood ratio would even be much, much larger.

\section{Discussion and conclusions}
\label{conclusions}
In accordance with the ENFSI guidelines, assessing evidence in court requires a Bayesian approach. This involves formulating at least two competing hypotheses, to be compared on the basis of evidence by a likelihood ratio. 

In a case of alleged cheating with secret codes, the prior assessment consists of two parts. Firstly, it is important to state a prior likelihood of cheating in general. In professional sports with high stakes, the probability of cheating will usually not be deemed astronomically small. Secondly, assuming a team cheats, a prior belief in an alleged code $C$ is called for. As we have seen in both the bridge and baseball case, the prior belief in $C$ can reasonably be bounded from below, because the way the codes have to be transmitted (orientation of cards, banging trash cans) do not allow for too many variations given that they should be useful. 

To calculate a likelihood ratio, one needs to set up models for both competing hypotheses. This may require some careful assumptions, but often the particular details will ultimately not really matter. Both the bridge and baseball cases are good examples of this, since simple but reasonable models led to astronomical likelihood ratios in favor of cheating with $C$ and $B$ respectively versus no cheating. We note that {\em any} statistical approach would need such modeling.

We further note that all other statistical reports on these cases do not fully incorporate Bayesian reasoning. The reason is usually that the Bayesian approach requires incorporating prior beliefs, which are often disregarded as `subjective'. As stated earlier, this aspect should not be viewed as a weak element of this framework, but rather as a reflection of reality.

Indeed, while prior beliefs may be personal, they are not formed arbitrarily. In this case, the lower bounds that we used for the priors are a product of common sense combined with sport-specific expertise. They may vary among different experts and judges, but in the light of overwhelming evidence these nuances will not be important. This demonstrates that this part of the analysis is neither automatically weak nor necessarily controversial. On the contrary, we view the prior assessment as an advantage of the Bayesian approach, because it allows for processing important contextual information that would otherwise have been ignored.

Although this paper focused on cheating with codes, the fundamental ideas should also work for other integrity issues in sports. In any such case, one will have to provide the two key ingredients of this Bayesian framework, that is, at least two competing hypotheses, accompanied by suitable models that allow a computation of the likelihood of the evidence. Thus, our contribution is to demonstrate how this framework can be developed for a specific type of alleged cheating.

In other situations, the principle is the same but the models are different. For example in chess, there are cases of suspicious behavior involving frequent toilet use together with exceptional performance at the chess board \cite{doggers}, giving rise to the hypothesis of cheating by checking a chess engine through a cell phone. In this case, the models need to involve two aspects: the quality of the chess moves and the frequency of the toilet use. There exist models that measure the performance based on the moves played \cite{haworth}, so that overperformance can reasonably be detected. For the toilet use, there are probably various reasonable choices, e.g.\ a Poisson model, under which calculations can be performed. Naturally, under the hypothesis of cheating with an engine, frequent toilet use together with a high performance becomes better explainable by a cheating than by an innocence hypothesis.

Another example is alleged use of doping without a positive test. If we return to the example of the women's 3000m track event from the introduction, it could be the case that an athlete failed to correctly inform the authorities about her whereabouts, resulting in missed doping controls. As in the previous example, provided that suitable models apply, this framework should yield a likelihood ratio to evaluate these events in the light of two competing hypotheses.

\section*{Acknowledgments}

The authors would like to thank the reviewer for their constructive comments and suggestions. In addition, we thank Kenneth Frank, Tom van Overbeeke, Emile Schols and Aldert Westra Hoekzema for their helpful comments.

\bibliographystyle{plain}

\bibliography{References}

\begin{appendices}

\section{Impact of cheating with leads}
\label{uitleg}

Recall the fact that if a pair plays a contract, one member becomes dummy, revealing their hand to both their partner and the opponents for the rest of the board. The partner of the dummy becomes declarer and is in full control of both hands.

The revealing of the dummy's hand has large implications. 
The other three players can now see both their own cards and the dummy's cards.  That means that still 26 cards are hidden, but all these 26 cards are divided among just two players. So if you're defending a contract and able to deduce that a specific card, say the ace of spades, is not with partner, you immediately know that declarer must have this card.
More generally, observe that if one of the three hidden hands is revealed as well, the board becomes an open book for the two remaining players.  Playing with open cards is therefore often called playing {\em double dummy}. For example, when a computer analyses a specific board based on knowledge of the position of all 52 cards, this is called a {\em double dummy analysis}. 

For professional bridge players, it is usually straightforward to find the optimal line of play in a double dummy scenario. Other than in chess, there is no individual time control, allowing players to take long thinks if needed. For that reason, computers do not perform significantly better in a double dummy setting, which may seem strange for people acquainted with chess.
The best players distinguish themselves through excellent deductions on the closed hands. Very often, top players are able to place the important cards by combining many small pieces of information. Alternatively, they may reduce all uncertainty to a limited number of likely scenarios, yielding incidental mistakes but a very good performance in the long run. 
When an average  chess player plays a grandmaster he often has the impression of playing against a computer. However, when you play against a pair of bridge masters, it feels as if your opponents have secretly checked your cards before the game.

The challenges of declaring or defending a bridge hand are similar to solving a logical puzzle. At the bridge table, putting all clues together may lead to a `solution' in the sense that all important unknown cards can be allocated. Much more often however, the puzzle cannot be completely solved. Instead, professional players are able to reduce their uncertainty to a small number of possible scenarios. One can imagine that one tiny extra clue can very well cause a cascade of new deductions, resulting in significantly better decisions.
For a non-bridge player some of the examples of illegal extra information may seem very minor, but the impact can easily be underestimated.

\section{Meaningful codes}
\label{codes}
Theoretically, there are many possible codes that transmit information about a hand by leading either horizontally or vertically. Given a specific sequence of boards combined with an oriented lead, there probably exist codes that fit these data perfectly. These could be very sophisticated codes, using all kinds of special conditions and consequences, but it could also be a very simple code.

In our analysis, we focused on {\em meaningful H-V codes}. A meaningful H-V code is a framework of signals to transmit information within a partnership with the following characteristics:

\begin{enumerate}
\item The code provides a tangible advantage to the players;
\item The code can be realistically implemented;
\item The information is transmitted by leading either horizontally or vertically, that is, in a binary way.
\end{enumerate}

Here are some examples of codes.
\begin{example}
 If a card is led horizontally, this implies that the leader holds the deuce of clubs. If the card is lead vertically, this denies the holding of this card.
\end{example}
This simple code provides information about the location of a single, often unimportant card. If instead the ace of clubs was the specific card, the code would be more useful. Only in very unlikely scenarios, this information about the deuce may be a deciding factor. This code fails to satisfy the first characteristic and is not considered a meaningful H-V code.

\begin{example}
\label{exam2}
 If a card is led horizontally while the board number is even (every board is numbered), this implies an ace in that suit together with at least one queen in some other suit. If the board number is odd, a horizontal card promises either the king or the jack in that suit. A vertical card on an even board number denies the ace unless the leader has no queens. In case of an odd board number, a vertical card implies either holding both king and jack or neither of these two cards.
\end{example}
Although this code may sometimes be useful, it is unlikely that it would be used by anyone. Alternatively, one could just use the orientation of the lead to imply or deny the ace in suit. This is probably just as effective, while also being much simpler.Once illegal signaling is agreed upon, such a code would most likely be used instead of the complicated example, since that one is so much harder to implement. This is an example of a code that could be constructed to fit certain data.

\begin{example}
In the FN-case, the alleged code, denoted $C$, is an example of a meaningful H-V code. It reads as follows. If a card is led horizontally, this denies a top honor (ace, king or queen) in the lead suit. Conversely, if the card is lead vertically, this implies holding a top honor in that suit. This rule does (naturally) not apply in case the lead card was a singleton, that is, the lead card was the only card in that suit.
\end{example}
    
\end{appendices}
\end{document}